\documentclass[aps,prb,twocolumn,showpacs,superscriptaddress]{revtex4}
\usepackage{epsfig}

\def\be{\begin{equation}} \def\ee{\end{equation}}
\def\bea{\begin{eqnarray}} \def\eea{\end{eqnarray}}

\begin{document}
\title{Topological invariants for interacting topological insulators with inversion symmetry}

\author{Zhong Wang$^1$, Xiao-Liang Qi$^2$ and Shou-Cheng Zhang}

\address{
Institute for Advanced Study, Tsinghua University, Beijing,  China, 100084}

\address{Department of Physics, Stanford University, CA 94305}

\date{\today}

\begin{abstract}

For interacting ${\rm Z}_2$ topological insulators with inversion symmetry, we propose a simple topological invariant expressed in terms of the parity eigenvalues of the interacting Green's function at time-reversal invariant momenta and zero frequency. We derive this result from our previous formula involving the integral over the frequency-momenta space. This formula greatly simplifies the explicit calculation
of ${\rm Z}_2$ topological invariants in inversion symmetric insulators with strong interactions.

\end{abstract}


\maketitle

\section{Introduction}

Topological insulators are quantum states of matter with insulating bulk and stable metallic surface\cite{qi2010a, moore2010, hasan2010,qi2011}. The origin of their interesting physics is the bulk topology in momentum space, which can generally be characterized by topological invariants, the simplest examples among which are the TKNN invariant\cite{thouless1982} and the ${\rm Z}_2$ topological invariants\cite{kane2005b,moore2007,Roy2009,qi2008,wang2010a}. The topological band invariants are defined in terms of the Bloch states, therefore, strictly speaking they can apply only to non-interacting systems, although one may extrapolate them to weakly interacting systems. On the other hand, the axion angle in the electromagnetic response can provide a general definition of an interacting topological insulator\cite{qi2008}.
In recent years, after the prediction and discovery of weakly interacting topological insulators, topological insulators with strong interaction have been among the main topics in the field\cite{raghu2008,shitade2009,zhang2009b,seradjeh2009,pesin2010,fidkowski2010,
li2010,dzero2010,rachel2010,zhang2011}. It is therefore urgent to search for topological invariants whose applicability is beyond the the non-interacting limit. One of the promising directions is the Green's function approach\cite{wang2010b}, which evaluates the axion angle and remains valid in the presence of interaction, provided that we use the full single particle Green's function. Much recent interest has been focused on this approach\cite{wang2011,wang2011a,gurarie2011,chen2011}.

Motivated by the topological field theory approach and the idea of dimensional reduction\cite{qi2008,qi2011},
we previously proposed a ${\rm Z}_2$ topological invariant in terms of Green's function
in the extended frequency-momentum space\cite{wang2010b}:
\bea
P_{3} &=& \frac{\pi}{6} \int_{0}^{1}du \int
\frac{d^{4}p}{(2\pi)^{4}} \textrm{Tr} [\epsilon^{\mu \nu \rho \sigma}
G\partial_{\mu}G^{-1} G\partial_{\nu}G^{-1} \nonumber
\\  &\,& \times G\partial_{\rho}G^{-1} G\partial_{\sigma}G^{-1}
G\partial_{u}G^{-1}]  \label{p3} \eea
in which the Greek letters are four-dimensional momentum-frequency indices including $p_0=i\omega$, and $u$ is an extension parameter between the physical Green functions at $u=0$ and a trivial constant reference function at $u=1$. For different extrapolations the value of $P_3$ obtained may differ by integer, but the fractional part of $P_3$ mod $1$ is independent from the extrapolation and is a physical topological invariant. This topoological invariant has the form of
a Wess-Zumino-Witten (WZW) term, which is not easy to implement in practical calculation.
In the present paper, we propose much simpler form of topological invariants [Eq.(\ref{parity}) and Eq.(\ref{2dparity})] for interacting topological insulators with inversion symmetry. In this new topological invariant, Green's function at eight time reversal invariant momenta (TRIM) $\vec{\Gamma}_i$ is evaluated and there is no integral involved, which greatly simplifies the calculation. This topological invariant can be regarded as a highly nontrivial generalization of the Fu-Kane invariant\cite{fu2007a} for non-interacting system. From the mathematical perspective, the logic leading to our formula is analogous to that of ref.\cite{wang2010a}.

\section{Topological invariant for three-dimensional topological insulators via Green's function eigenvectors at zero frequency}
In this paper we focus on three-dimensional time-reversal invariant insulators with inversion symmetry. The time reversal operator is an anti-unitary operator defined by $\hat{T} = T\hat{K}$, where $\hat{K}$ is the complex conjugation operator and $T$ is a matrix satisfying $T^* T=-1$. The inversion transformation matrix $P$ satisfies $P^2=1$, therefore, its eigenvalues are $\pm 1$.
The Green's function is a $N\times N$ matrix, where $N$ is an even integer due to time reversal symmetry.  It is convenient for our purpose to use the imaginary frequency Green's function, e.g. a free fermion system with Hamiltonian $H= \sum_k c^\dag_k h(k) c_k$ has Green's function $G(i\omega,k)=1/[i\omega-h(k)]$. Generally, we have the Schwinger-Dyson equation $G^{-1}(i\omega,k)=G^{-1}_0 (i\omega,k)-\Sigma(i\omega,k)$ between the full single particle Green's function $G$, the non-interacting Green's function $G_0$, and the self energy $\Sigma$. A general $G(i\omega,k)\in {\rm GL}(N,C)$ can be diagonalized to give the eigenvalues and eigenvectors.
Explicitly, we have
\bea G^{-1}(i\omega,k)|\alpha(i\omega,k)\rangle &=& \mu_\alpha (i\omega,k)|\alpha(i\omega,k)\rangle  \label{eigin} \eea
where $\mu_\alpha(i\omega,k) \equiv a_\alpha(i\omega,k)+i b_\alpha(i\omega,k)$, with $a_\alpha, b_\alpha$ being real number.
The matrix $G(i\omega,k)$ has the same eigenvectors but eigenvalues $\mu_\alpha^{-1}(i\omega,k)$.
Because of inversion symmetry, Green's function satisfies $PG(i\omega,k)=G(i\omega,-k)P$. At eight TRIMs $k=\vec{\Gamma}_i$, where $\vec{\Gamma}_i$ is equivalent to $-\vec{\Gamma}_i$, the Green's function satisfies $PG(i\omega,\vec{\Gamma}_i)=G(i\omega,\vec{\Gamma}_i)P$, therefore, $|\alpha\rangle$ can be chosen as the simultaneous eigenvectors of $G$ and $P$, namely that in addition to eq.(\ref{eigin}), $|\alpha\rangle$ also satisfies the equation \bea
P|\alpha(i\omega,\vec{\Gamma}_i)\rangle =\eta_\alpha |\alpha(i\omega,\vec{\Gamma}_i)\rangle \eea where $\eta_\alpha$ is the parity eigenvalue.
Using the Lehmann representation, we can obtain the equation \bea G^\dag(-i\omega,k)=G(i\omega,k) \label{hermitian}\eea from which it follows that $G(0,\vec{\Gamma}_i)$ is a Hermitian matrix and thus has real eigenvalues, i.e. $\mu_\alpha(0,\vec{\Gamma}_i)$ is always located on the real axis of the $\mu_\alpha = a_\alpha+i b_\alpha$ complex plane. If $\mu_\alpha(0,\vec{\Gamma}_i)$ is located on the right half of the real axis, namely $\mu_\alpha(0,\vec{\Gamma}_i)>0$, we call $|\alpha(0,\vec{\Gamma}_i)\rangle$ a ``Right-zero (R-zero)''. Similarly we can define ``L-zero''.
One of the central results of this paper is the  topological invariant defined as
\bea (-1)^{\Delta} \equiv \prod_{ {\rm R-zero}  } \eta_\alpha^{1/2} =\pm 1 \label{parity} \eea
where we have used the convention that $(-1)^{1/2}=i$ for the square root. The appearance of square root can be interpreted as follows. From the equations $T^{-1}G^T(i\omega,\vec{\Gamma}_i) T =G(i\omega,\vec{\Gamma}_i)$\cite{wang2010b,gurarie2011} (the superscript $^T$ is matrix transposition) and Eq.(\ref{hermitian}),  we can show that \bea G^{-1}(-i\omega,\vec{\Gamma}_i)T|\alpha(i\omega,\vec{\Gamma}_i)\rangle^* =\mu_\alpha^* (i\omega,\vec{\Gamma}_i))T|\alpha(i\omega,\vec{\Gamma}_i)\rangle^* \label{kramers} \eea Therefore, R-zeros $|\alpha(0,\vec{\Gamma}_i)\rangle$ and $T|\alpha(0,\vec{\Gamma}_i)\rangle^*$ always form a Kramers pair. The square root in eq.(\ref{parity}) amounts to requiring that each Kramers pair with $\eta_\alpha=-1$ contributes $i^2=-1$. We will provide a derivation of eq.(\ref{parity}) in the following sections, and also present an equivalent formula in eq.(\ref{curve}).

In the non-interacting limit, the parity of Green's function eigenvectors is reduced to the parity of Bloch states, and it can thus be
shown that our formula in eq.(\ref{parity}) is reduced to the Fu-Kane formula\cite{fu2007a}.

\section{An equivalent formula and the topological invariance}
\begin{figure}
\includegraphics[width=8.0cm, height=9.5cm]{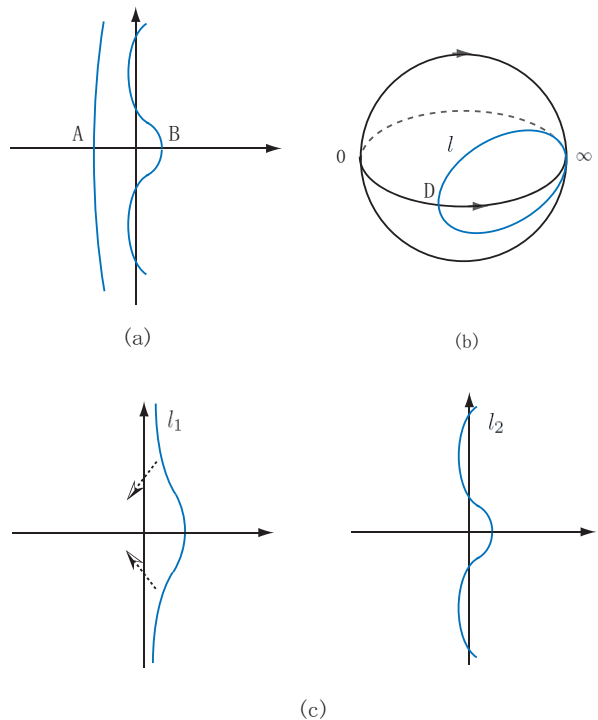}
\caption{Topological invariance of R/L-curves on the $\mu_\alpha$ complex plane. (a) $A$ and $B$ is L-zero and R-zero respectively, and the corresponding curve is L/R-curve respectively. (b) Compactified complex plane, on which a R-curve $l$ is shown.  Here the only crossing point of $l$ with real axis is $D$. The point $\infty$ is not reached by $l$, although $l$ approaches $\infty$ as $\omega\rightarrow\pm\infty$. (c) The R-curve $l_1$ can be deformed into another R-curve $l_2$ following the path indicated by the arrow, however, it cannot be deformed into a L-curve if $0$ and $\infty$ are avoided, i.e. topological class is preserved during smooth deformations. }
\label{class}
\end{figure}

Let us begin with some notations.
As $\omega$ flows from  $-\infty$ to $+\infty$,  $\mu_\alpha (i\omega,\vec{\Gamma}_i)$ sweeps a curve on the complex plane.
There are two special points $0$ and $\infty$ on this complex plane, which can be compactified by adding a point $\infty$ at infinity. We assume that $G(i\omega, \vec{\Gamma}_i) \in {\rm GL}(N,C)$, therefore, $\mu_\alpha \neq 0 \, {\rm or}\, \infty$. Because $\mu_\alpha(i\omega,k)\rightarrow i\omega$ as $\omega \rightarrow \pm \infty$, each $\mu_\alpha$ curve must cross the real axis by an odd number of times. All $\mu_\alpha$ curves can be classified into two topological classes: ``right $\mu_\alpha$ curves'' (R-curves) and ``left  $\mu_\alpha$ curves'' (L-curves). The R/L-curves, by our definition, cross the right/left half real axis at $i\omega=0$[see Fig.\ref{class}a]. It is also worth noting that  the $\mu_\alpha$ curve cross the real axis only at $i\omega=0$, which follows from the Lehmann representation. Due to Eq.(\ref{hermitian}), each R/L-curve is symmetric according to the real axis [see Fig.(\ref{class}a)].  Furthermore, from Eq.(\ref{kramers}) we see that each R/L-curve has a Kramers partner which coincides with it. It should be emphasized that the distinction between R-curves and L-curves is topological because an R-curve cannot be smoothly deformed to a L-curve without crossing the two singular points $0$ and $\infty$, nor \emph{vice versa} [see Fig.(\ref{class})].

Now we can define a new topological invariant in terms of R/L-curves as
\bea (-1)^{\Delta} = \prod_{{\rm R-curve}} \eta_\alpha^{1/2} =\pm 1 \label{curve} \eea
which is equivalent to eq.(\ref{parity}), as we will see.
The parity formulas in eq.(\ref{parity}) and eq.(\ref{curve}) are the central results of this paper.  The topological invariance of eq.(\ref{curve}) follow from the topological invariance of  R/L-curves and the fact that the parity $\eta_\alpha$, which takes only two discrete values $\pm 1$, cannot changes abruptly during a smooth deformation. It is clear that eq.(\ref{parity}) is more efficient in practical calculation, while eq.(\ref{curve}) provides a deeper perspective, for instance, we note that the definition of ``R/L-curve'' can be modified as follows: We can arbitrarily choose a straight line (other than the imaginary axis ) through the singularity $0$, and define ``R/L-curve'' according to the locations where $\mu_\alpha$ curves cross this line. It can be easily appreciated that the topological class of an R/L-curve is independent of the choices of this straight line. Eq.(\ref{curve}) reveals such degree of freedom in the definition of R/L-curve.

We have assumed that Green's function has no singularity, which is a mild assumption for an insulator. However, if we want to study phase transitions, singularities must be considered.
There are two types of singularity of Green's function, one of which is the crossing of $\mu_\alpha$ curves with $0$, the other is crossing with $\infty$. Topological phase transitions must be accompanied by these singularities. The transitions of first type (singularity at $0$)  include all topological phase transitions in free fermion systems, while those of the second type (singularity at $\infty$) can happen when the interaction is sufficiently strong.

\section{Derivation of the new formula from the previous P$_3$ formula}
In this section we provide a derivation of the parity formula in eq.(\ref{parity}) and eq.(\ref{curve}).   It was shown in ref.\cite{wang2010b} that the ${\rm Z}_2$ topological invariant of interacting topological insulators can be expressed in terms of Green's function as in Eq. (\ref{p3}). It can be expressed
more compactly as
\bea 2P_{3} &=& W(G)|_{R\times T^4} \nonumber \\
&\equiv & \frac{1}{480\pi^3} \int_{-\pi}^{\pi}d^5p \textrm{Tr} [ \epsilon^{\mu \nu \rho \sigma \tau}
G\partial_{\mu}G^{-1}  \nonumber
\\  &\,& \times  G\partial_{\nu}G^{-1} G\partial_{\rho}G^{-1} G\partial_{\sigma}G^{-1}
G\partial_{\tau}G^{-1}] \nonumber \\ &=& {\rm integer} \label{wzw} \eea
where $W(G)|_{R\times T^4}$ is the ``winding number'' of the map from frequency-momenta space $R\times T^4$ to $GL(N,C)$.  We have used the conventions that $p=(p_0,p_1,p_2,p_3,p_4)$, $p_0\equiv i\omega$ is the frequency, $(p_1,p_2,p_3)$ are momenta, and $p_4$ is the WZW parameter. $(p_0,p_1,p_2,p_3,0)$ expands the physical frequency-momenta space, and $(p_0,p_1,p_2,p_3,\pi)$ expands the frequency-momenta space of the trivial reference system.
We have assumed that the eigenvalues of Green's function is everywhere smooth and non-vanishing. The topological classes of ${\rm Z}_2$ insulators are defined by $2P_3$ mod $2$. For inversion symmetric system, we can also add the constraint that $PG(p_0,p_1,p_2,p_3,p_4)=G(p_0,-p_1,-p_2,-p_3,-p_4)P$.

With this preparation, we would like to obtain the equation $2P_3=\Delta\,{\rm mod}\,2$, in which $2P_3$ is defined by eq.(\ref{wzw}) and $\Delta$ is defined by eq.(\ref{parity}) or eq.(\ref{curve}). The basic idea is outlined as follows. Both $2P_3$ and  $\Delta$ are topologically invariant, therefore, as we deform $G(p)$ as a $N\times N$ matrix, $2P_3$ and $\Delta$ can only change abruptly when $G(p)$ encounters singularities. We can compute their changes, and we will find that they are equal mod $2$. Now we only need one example in which $2P_3=\Delta\,{\rm mod}\,2$ to prove this identity in general cases, but we already know that $2P_3=\Delta\,{\rm mod}\,2$ in the non-interacting limit\cite{wang2010a,wang2010b}, therefore, $2P_3=\Delta\,{\rm mod}\,2$ is generally correct.

Now we present our calculations following the above outline. When we talk about deformation of $G(p)$, the precise meaning is that $G=G(p,\lambda)$, where $\lambda$ parametrizes the deformation. Without losing generality, we assume the singularity occurs at $\lambda=0$. We consider a singularity located at $p=(0,0,0,0,0)$ for the following reasons (for the spatial components, $p_i=0; \,i=1,2,3$ means the TRIM). The spatial components $(p_1,p_2,p_3)$ are assumed to be one of the eight TRIMs, because singularities away from TRIMs must occur pairwise at  $(p_1,p_2,p_3,p_4)$ and $(-p_1,-p_2,-p_3,-p_4)$ due to inversion symmetry, which can thus only change the value of $W(G)|_{R\times T^4}$ by an even integer. Similarly, the frequency $p_0=0$ because of the equation $G^\dag(-p_0,\vec{\Gamma}_i)=G(p_0,\vec{\Gamma}_i)$, which requires that the singularities occur pairwise at $p_0$ and $-p_0$.

First we consider the singularity defined by ${\rm det}\left[G^{-1}(p,\lambda=0)\right]=0$. We can expand $G^{-1}(p,\lambda)$ in a neighborhood keeping only terms linear in $p$. In the cases when the first order terms vanish and higher order terms dominate the expansion, we can always perturb the singularity to split a single singularity into several linear ones. At a generic linear singularity, four eigenvalues of $G^{-1}(p,\lambda=0)$ vanish, as is required by time-reversal and inversion symmetries. Therefore it is sufficient to write only the singular part of the Green's function, which has the following form:
\bea G^{-1}(p,\lambda) = u_0(p_0,\lambda) +\sum_{\alpha=1}^4 u_\alpha(p_i) \Gamma_\alpha  + u_5(p_0,\lambda)P \label{expansion} \eea
where  $u_0(p_0,\lambda)=C_{00} p_0 + C_{05} \lambda$, $u_5(p_0,\lambda)=C_{50} p_0 + C_{55} \lambda$, and $u_\alpha =\sum_{i=1}^4 C_{\alpha i}p_i \,(\alpha=1,2,3,4)$. Generically we have $\det C\neq 0$.
The matrices appearing here are $P=1\otimes\tau_3$, $\Gamma_\alpha=\sigma_\alpha\otimes \tau_1 (\alpha=1,2,3)$, $\Gamma_4=1\otimes\tau_2$, where both $\sigma$ and $\tau$ are Pauli matrices. Time reversal matrix is defined as $T=i\sigma_2\otimes 1$ in our notation. The form of this expansion is dictated by $PG(p_0,p_i)=G(p_0,-p_i)P$ and $T^{-1}G^T(p_0,p_i) T =G(p_0,-p_i)$, e.g. inversion symmetry requires that $\Gamma_\alpha \,(\alpha=1,2,3,4)$ contain factor $\tau_1$ or $\tau_2$ but not $\tau_3$.
We note that the expansion in eq.(\ref{expansion}) is analogous to that in ref.\cite{murakami2007,murakami2008}, in which an expansion of Hamiltonian according to symmetries was used.

\begin{figure}
\includegraphics[width=9.0cm, height=5.0cm]{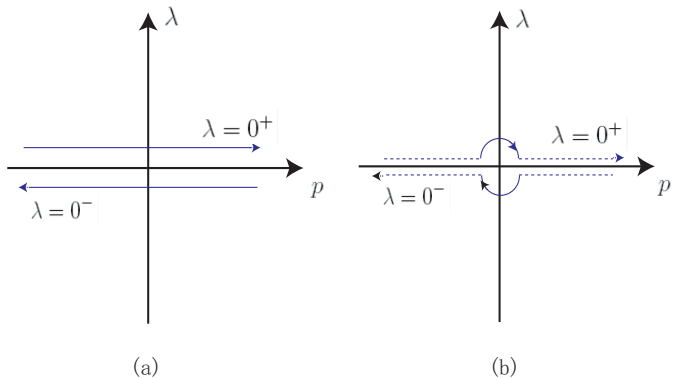}
\caption{The deformation of integral contours. The  $p$ axis refers to $p=(p_0,p_1,p_2,p_2,p_4)$. The two contours in (a) correspond to $\lambda>0$ and $\lambda<0$ respectively.  The opposite orientations come from the negative sign in $-W(G,\lambda=0^-)$. We can deform these two contours to those illustrated in (b), which does not change $W(G)$ due to topological invariance.
In the  calculation of the jump of $W(G)$, the ``dashed lines''(which are in fact 5d manifolds) in (b) cancel each other, and we are left with the ``solid circle'', which is the five dimensional sphere $S^5$. }
\label{s5}
\end{figure}

We now proceed to calculate the change $W(G,\lambda=0^+)|_{R\times T^4}-W(G,\lambda=0^-)|_{R\times T^4}$. To this end, we define a five dimensional sphere $S^5$  by $\sum_{\mu=0}^4 |p_\mu|^2 +\lambda^2=r^2$, whose radius $r$ is sufficiently small, then we have a mapping $f$ from $S^5$ to the manifold parameterized by $\{(u_0,u_1,u_2,u_3,u_4,u_5)\}$. From a simple geometrical picture[see Fig.\ref{s5}], we have $W(G,\lambda=0^+)|_{R\times T^4} - W(G,\lambda=0^-)|_{R\times T^4} = W(G)|_{S^5}$.
For the standard mapping $(p_0,p_1,p_2,p_3,p_4,\lambda)\rightarrow (u_0,u_1,u_2,u_3,u_4,u_5)=\frac{R}{r}(p_0,p_1,p_2,p_3,p_4,\lambda)$, where $R$ is a nonzero constant, we can obtain $W(G)|_{S^5}=1$ by strightfoward calculations. Because of the linear relation between $u_\alpha \,(\alpha=0,1,2,3,4,5)$ and $p_0,p_1,p_2,p_3,p_4,\lambda$, we can obtain the general result $W(G)|_{S^5}=\pm 1$.

At the same time, we can see that the R/L-zeros have $a_\alpha = {\rm Re}(C_{05}+C_{55}\eta_\alpha)\lambda, \, \eta_\alpha=\pm 1$, both of which are two-fold degenerate. We look at the two eigenvectors with $\eta_\alpha=-1$ because only those contribute nontrivially to $\Delta$. When $\lambda$ changes sign, the two R/L-zeros with $\eta_\alpha=-1$ will change their topological classes, i.e. from being R-zeros to being L-zeros or \emph{vice versa}.  Correspondingly, two R/L-curves with $\eta_\alpha=-1$ change their topological classes.  This will generate a  factor $i^2=-1$ in $(-1)^\Delta$. Therefore, we see that the changes of $2P_3$ and $\Delta$ match each other. In the above analysis we consider an singularity with vanishing ${\rm det}\left[G^{-1}(p)\right]$, but we can also consider singularities with vanishing ${\rm det}\left[G(p)\right]$. The linear expansion around the singularity is the same as that in vanishing ${\rm det}\left[G^{-1}(p)\right]$ cases.
In summary, we show that for every linear singularity,  the changes of $2P_3$ and $\Delta$ are always equal mod $2$. Since all other types of singularity can generically be perturbed to linear ones, we have proved our statement that $2P_3=\Delta\,{\rm mod}\,2$.

By an analogous calculation, one can show that a four (spatial) dimensional time reversal invariant insulators with inversion symmetry is topologically nontrivial if $\prod_{ {\rm R-zero}  } \eta_\alpha^{1/2}=-1$. In this parity approach, only the ${\rm Z}_2$ part of ${\rm Z}$ classification can be detected. When the four-dimensional topological insulator is dimensionally reduced to three dimensions, the robust topological invariant is exactly the ${\rm Z}_2$ part.

\section{Interacting topological invariants for two-dimensional topological insulator with inversion symmetry}

For two dimensional time reversal invariant and inversion symmetric topological insulators, similar formulas like eq.(\ref{parity}) and eq.(\ref{curve}) can also be defined, with the only difference that there are now four instead of eight $\vec{\Gamma}_i$ points. Explicitly, we have

\bea (-1)^{\Delta_{2D}} = \prod_{{\rm R-zero}} \eta_\alpha^{1/2} = \prod_{{\rm R-curve}} \eta_\alpha^{1/2} \label{2dparity} \eea
whose topological invariance and derivation follows the same logic as the previous section, which we shall not repeat.

\section{Conclusions}

In this paper we proposed a simple form of the ${\rm Z_2}$ topological invariant for interacting topological insulators with inversion symmetry.  These formulae are well defined provided that the Green's function has no singularities. In these cases, our formulae provide a ${\rm Z}_2$ classification of interacting topological insulators.  It should be mentioned that this is not a complete classification of time reversal invariant insulators because those states with ground state degeneracy are not included, as mentioned in Ref.\cite{wang2010b}. However, most of the realistic models and materials with inversion symmetry are within the description presented in this paper, thus the new formulae can greatly simplify the evaluation of ${\rm Z}_2$ topological invariants.

In the inversion-asymmetric case, if one can tune physical parameters so that the system smoothly evolves into a inversion-symmetric one, i.e. the Green's function is nonsingular during the evolution, then the ${\rm Z}_2$ class of the original system can be inferred from that of the inversion-symmetric system, which can be calculated using our parity formula. This parity formula can be calculated explicitly using various numerical methods.  It is also worth mentioning that our formula is also applicable to the so-called ``inversion symmetric topological insulator'', i.e. topological insulators protected only by inversion symmetry\cite{hughes2011,turner2010a,wan2011}.

ZW acknowledges financial support (No. 553401001) from Tsinghua University. XLQ is supported by the Packard Foundation. SCZ is supported by the NSF under grant numbers DMR-0904264 and the Keck Foundation.

\bibliography{parity}

\end{document}